\documentclass[usenatbib,usegraphicx]{mn2e}
\usepackage{amsmath}
\newcommand{\eg}{{\it e.g.,}~}
\newcommand{\ie}{{\it i.e.,}~}
\newcommand{\Msun}{M_{\odot}}
\newcommand{\hMsun}{~h^{-1}{\rm M}_{\odot}}

\newcommand{\LCDM}{$\Lambda$CDM~}
\newcommand{\beq}{\begin{equation}}
\newcommand{\eeq}{\end{equation}}
\newcommand{\Mstar}{M_{\star}}

\newcommand{\fihl}{f_{\mathrm{IHL}}}
\newcommand{\ohihl}{[O/H]_{\mathrm{IHL}}}

\begin{document}

\title{The Metallicity of Diffuse Intrahalo Light}

\author[C. W. Purcell et al.]
{Chris W. Purcell,$^1$
James S. Bullock,$^1$
and Andrew R. Zentner$^2$ \\
$^1$Center for Cosmology, Department of Physics and Astronomy, The University of California, Irvine, CA 92697 USA \\
$^2$Department of Physics and Astronomy, University of Pittsburgh, Pittsburgh, PA 15260 USA
}

\date{Accepted 2008 September 8.  Received 2008 September 4; in original form 2008 May 19}

\maketitle

\begin{abstract}  

We make predictions for the metallicity of diffuse 
stellar components in systems ranging from small spiral galaxies to
rich galaxy clusters.  We extend  the formalism of
\citet{Purcell_etal07}, in which diffuse stellar mass is produced via
galaxy disruption, and we convolve this result  with the observed
mass-metallicity relation for galaxies in order to analyze the
chemical abundance of intrahalo light (IHL) in host halos with virial
mass $10^{10.5} \Msun \leq M_{host}  \leq 10^{15} \Msun$.  We predict
a steep rise of roughly two dex in IHL metallicity from the scales of
small to large spiral galaxies.  In terms of the total dynamical mass $M_{host}$ 
of the host systems under consideration, we predict diffuse 
light metallicities ranging from $Z_{IHL} \la -2.5$ for $M_{host} \sim
10^{11} \Msun$,  to $Z_{IHL} \sim -1.0$ for $M_{host} \sim 10^{12}
\Msun$.  In larger systems, we predict a more shallow rise in 
this trend with $Z_{IHL} \sim -0.4$ for $M_{host}  \sim 10^{13} \Msun$, 
increasing to $Z_{IHL} \sim 0.1$ for $M_{host} \sim 10^{15} \Msun$.   
This behavior is coincident with a narrowing of the intrahalo
metallicity distribution as host mass increases.  The observable
distinction in surface brightness between old, metal-poor IHL stars
and more metal-rich, dynamically-younger tidal streams is of crucial
importance when estimating the chemical abundance of an intrahalo
population with multiple origins.

\end{abstract}

\begin{keywords}
Cosmology: theory -- galaxies: formation -- galaxies: evolution -- clusters: diffuse light
\end{keywords}

\section{Introduction}
\label{section:intro}

Diffuse stellar components are ubiquitous phenomena that include the
intracluster light  at the centers of rich galaxy clusters as well as
the diffuse stellar halos that surround individual  galaxies like the
Milky Way.  This diffuse luminosity comprises $\la 1-5$\% of the total
luminosity  of spiral galaxies
\citep{Morrison_etal00,Chiba_Beers00,Yanny_etal00,Ivezic_etal00,Siegel_etal02,Irwin05,
Guhathakurta_etal05,Kalirai06,Chapman_etal06} and as much as $\sim
10-40$\% of the luminosity of rich galaxy  clusters
\citep{cmbs00,Lin_Mohr04,Feldmeier_etal04,Mihos_etal05,Zibetti05,Krick_etal06,seigar06}.
Whether in clusters or surrounding individual galaxies, we refer to
this diffuse material as ``intra-halo light'' (IHL).   In the
prevailing theory of structure formation, objects are built
hierarchically through sequential mergers of smaller objects.   This
provides a natural means of IHL production via the disruption of
merging  galaxies
\citep{Gallagher_Ostriker72,Merritt83,Byrd_Valtonen90,Dubinski03,Gnedin03,
Mihos04,Murante_etal04,Lin_Mohr04}.  This mechanism includes the
emergence of diffuse stellar halos  about galaxies such as the Milky
Way, resulting from the merger and destruction of dim,  dwarf-sized
galaxies
\citep{Searle_Zinn78,Johnston_etal96,Johnston_etal98,Helmi_etal99,Bullock_etal01,jsb01,bj05,
Robertson_etal05,DeLuciaHelmi08}, making the IHL a potentially rich
testbed for structure and galaxy formation theories.   In a previous
paper, we made predictions for the relative prominence of the IHL as a
function of both  the total mass of the system (including luminous and
dark matter) and the luminosity of the brightest object  in each
system for systems ranging from dwarf galaxies to rich clusters
\citep{Purcell_etal07}.  In this  paper we seek to add to the
discriminatory power of the IHL as a test of galaxy formation models
by making  predictions for the metallicities of IHL components in such
systems.  

Observations of intracluster light in Abell 3888 indicate that the
stellar population composing it has a  metallicity distribution
function peaking near $Z \sim 1.0$ in the outer regions of the diffuse
component, and  $0.2 \le Z \le 0.5$ in the inner region
\citep{Krick_etal06}, although there is large uncertainty in both
values  stemming from chemical evolution systematics and color
analysis.  Conservatively, we can say that the  chemical abundance of
intrahalo stellar material in cluster-sized host systems is at least
one order of magnitude  greater than that of the diffuse stellar
components around large spirals.  M31's pressure-supported halo has an
approximate iron abundance of [Fe/H] $\sim -1.4$ \citep[][see also
\citealt{Kalirai06}]{Chapman_etal06}, and it  resembles the Milky Way
halo both kinematically and structurally.  The Milky Way's halo stars
are  also metal-poor at around [Fe/H]~$\sim -1.5$
\citep[\eg][]{ryan_norris91}.  However, M33's halo  has proven elusive
as there is only tentative evidence for a non-rotating, power-law
diffuse  component at roughly [Fe/H]~$\sim -1.5$ \citep[][see also
\citealt {McConnachie_etal06} for spectroscopic field
analysis]{Ferguson07}.

Numerical simulations indicate that the shredding of infalling
galaxies during the hierarchical  merging process of host halo
formation is a prevalent  contributor to diffuse light in clusters
\citep{Willman_etal04,Rudick_etal06,Sommer-Larsen06}.    There is
often sufficient stellar mass in this component to dominate the light
emitted by  the bright central galaxy in the cluster \citep[][see
\citealt{Gonzalez_etal05,Seigar_etal06} for  observational evidence
supporting this prediction]{Conroy_etal07}.   Likewise, simulations of
Galaxy formation support the accumulation of a spheroidal stellar halo
via the accretion and  disruption of dwarf-type satellite galaxies
during the construction of the primary system
\citep{Diemand_etal05,Read_etal06, Font_etal06,Abadi_etal06}.


\begin{figure}
\includegraphics[width=84mm]{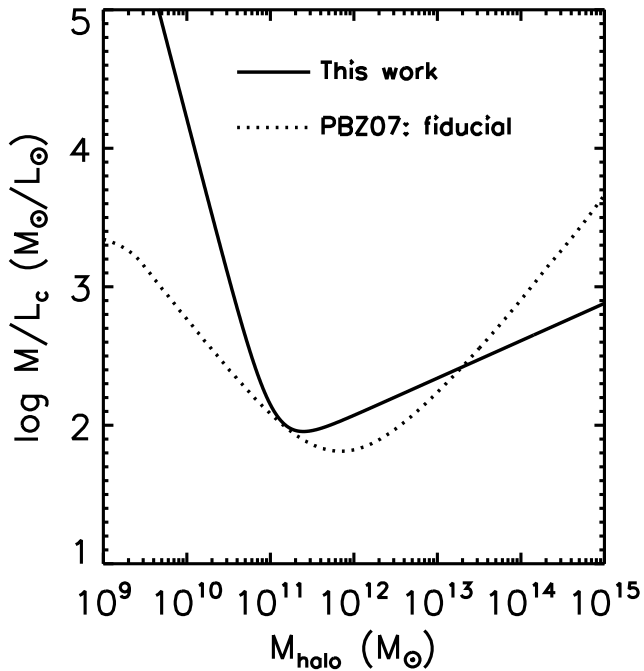}
\caption{
The global mass-to-light $M/L_c$ ratio as a function of host halo mass $M_{halo}$.  
The {\em solid} line shows the $M/L_c$ result obtained by \citet{vdb07} for 
the WMAP1 cosmological parameters.  For comparison with the model of PBZ07, 
the {\em dotted} line represents that work's fiducial $M/L_c$ as drawn from 
\citet{yang_etal03}.  
}
\label{fig:masslum}
\end{figure} 

Our aim in this paper is to predict the metallicity of intrahalo light
as a function of the total mass of the system.   We explore the
behavior of any trends in metallicity as a function  of
properties of disrupted progenitor systems, in order to discriminate
bright diffuse streams from dynamically-older  and more well-mixed
tidal debris.  This debris is created from an enormous dynamic range
in halo mass.   In addition, the system-to-system scatter in the
fraction of luminosity in IHL and IHL metallicities is significant,
indicating that large samples are needed in order to study 
IHL metallicity trends as a  function of mass.  These factors make
direct simulation of IHL production a computationally costly endeavor
that is  impractical for our purposes, particularly given the inherent
uncertainties in numerical treatments of baryonic processes.   As an
alternative, we rely on the analytic treatment prescribed by
\citet[][see below]{Zentner05} in order  to follow the coalescence of
small systems into larger composite systems.  This approach models the
merging rates and  subsequent interactions, including heating and
disruption, of systems in an approximate way but with the benefit of
being a computationally inexpensive approach that can be used to
minimize statistical (not systematic)  uncertainties in model
predictions.  We assume that the stellar material in a merging galaxy
will be diffused  into the intrahalo light of the larger system when
the galaxy's parent subhalo is significantly affected  by tidal and
heating processes.  Direct treatments of baryonic processes and star
formation are uncertain at  present.  As such, we employ a
``bootstrap'' approach that feeds the observed mass-metallicity
relation for galaxies into our calculation.  After
determining the stellar contents of disrupted, merging halos
according to the empirically-constrained model of \citet[][see
below]{Purcell_etal07}, we assign these stars a  metallicity based on
the observed mass-metallicity relation.  The metallicities of all
stars distributed throughout  the diffuse light by all merging and
disrupted systems are used to obtain the oxygen  abundance
$[O/H]_{\mathrm{IHL}}$ of diffuse intrahalo light as a function of the
total mass of the system $M_{host}$.  At this time, detailed and
robust predictions remain beyond the reach of theoretical modelling,
so we emphasize  \citep[as we did in our previous
study,][]{Purcell_etal07} that our goal is to highlight gross trends
that should  be expected in hierarchical models on very general
grounds.  

In the next section, we outline our model for IHL metallicities.   In
\S~\ref{sec:results}, we present our results  and discuss their
implications in the context of observational  benchmarks in chemical
abundance measurements of galactic  stellar halos as well as
intragroup/intracluster luminosity.   Throughout this work, we adopt a
$\Lambda$CDM cosmological  model with $h=0.7$,
$\Omega_m=1-\Omega_\Lambda=0.3$,  and a primordial power spectrum
which is scale-invariant,  $n=1$, and normalized to $\sigma_8=0.9$.

\section{Methods}
\label{sec:methods}

We investigate the metallicity of tidally-disrupted, diffuse stellar material as a function of 
host dark matter halo mass.  In our terminology, we refer to the largest dark matter halo 
surrounding a group or cluster of galaxies as the ``host'' dark matter halo and the smaller 
halos that are themselves contained within the host but self-bound systems as ``subhalos.''  
Our study expands upon the model of \citet[][hereafter PBZ07]{Purcell_etal07}, 
which makes statistical predictions for the stellar mass fraction expected 
in a diffuse, luminous intrahalo component, over a wide range of host 
halo mass ($10^{10.5} \Msun \leq M_{host} \leq 10^{15} \Msun$).  We give a brief outline of 
the relevant features of this model here.  A more complete discussion 
of our modeling techniques and definitions can be found in PBZ07.  

Out treatment utilizes the halo formation model of \citet[][for an earlier version see 
\citealt{zb:03}]{Zentner05} in order to track the accretion, evolution, and possible 
destruction of dark matter subhalos analytically.  
The model of \citet{Zentner05} produces mass accretion histories over a wide 
range in host halo mass according to the extended Press-Schechter (EPS) 
formalism \citep[][for a recent review see \citealt{zentner06}]{Bond91,LC93}, 
using the specific implementation described by \citet{Somerville99}, 
and then uses analytic prescriptions for dynamical friction and mass loss due to 
interactions within the larger host potential to evolve the population of accreted subhalos.  
This model is very well-tested against dissipationless cosmological simulations, 
reproducing a variety of subhalo statistics predicted by direct numerical techniques 
accurately.  We encourage the reader to refer to \citet{Zentner05} for more detail on this 
aspect of our model.  

Applying observational constraints from galaxy survey data, PBZ07 assign subhalo luminosities 
according to the results obtained by \citet[][see below]{yang_etal03} in their analysis of the 
conditional luminosity function of galaxies in the Two-Degree Field Galaxy Redshift Survey (2dFGRS).  
The model of PBZ07 then implements a critical mass-loss threshold at which orbiting subhalos are 
considered to be completely tidally shredded when their circular velocities fall below a critical fraction 
of their maximum circular velocity they had upon accretion into the host, $f_{\mathrm{crit}} = 
V_{\mathrm{crit}}/V_{max}(t_{acc})$.  A second free parameter $\alpha$ describes the star formation history 
of the subhalo, in order to determine its stellar-mass content upon accretion and the subsequent 
truncation of star formation via ram pressure stripping induced by the host halo's hot gas.  Both 
parameters are tuned such that contemporary populations of surviving galaxies are recovered 
by the fiducial model across the host mass scales of interest, and it is instructive to note that the results 
are insensitive to dramatic changes in these values, which are set to $f_{\mathrm{crit}} = 0.6$ and 
$\alpha = 1$.  We ignore contributions made to a system's diffuse light by surviving 
satellites, as simulations indicate that subhalos are typically disrupted shortly after their stars 
become tidally stripped in any substantial sense \citep{BullockJohnston05}.

The global variance in the mass-to-light ratio of dark halos as a function of host mass, as obtained 
by \citet{yang_etal03} and discussed in \S~\ref{subsec:upgrading}, convolved with the EPS formalism's 
prediction that host halos are chiefly built by the accretion of subhalos roughly a tenth as massive, yield 
a clear correlation between the relative amount of intrahalo light in a system and the total mass of the 
system.  This predicted correlation spans three orders of magnitude in mass from galactic scales ($\fihl 
\sim 1-5\%$ for $M_{host} \sim 10^{12} \Msun$) to those of galaxy clusters ($\fihl \sim 10-40\%$ for 
$M_{host} \sim 10^{15} \Msun$), and we refer the reader to \citet{Purcell_etal07} for a complete 
description of the method and prior results.

\subsection{Intrahalo Light Model Update}
\label{subsec:upgrading}

An ingredient leading to the correlation between IHL fraction and host 
mass is the aforementioned variation of mass-to-light ratio ($M/L_c$) 
with halo mass, which is necessary in the prevailing structure formation 
model in order to reproduce empirical measurements 
of the galaxy luminosity function and clustering statistics 
\citep[\eg][]{White78,kwg93,sp:99,tinker:05,cm:05}.  PBZ07 adopt the particular 
result of \citet{yang_etal03}, in which the authors constrain 
the conditional luminosity function (CLF) of galaxies in the Two-Degree Field Galaxy 
Redshift Survey (2dFGRS) and infer the characteristic B-band luminosity, 
$L_c(M)$, for the brightest central galaxy sitting in a host halo of virial 
mass $M$.  In a new analysis by the same group, \citet{vdb07} investigate 
variation in their CLF results as a function of the cosmological parameters 
measured by WMAP~\footnote{In this paper, we will refer to three cosmologies 
corresponding to WMAP data releases: WMAP1 with $\Omega_m=0.3$ and 
$\sigma_8=0.9$, WMAP3 with $\Omega_m=0.23$ and $\sigma_8=0.74$, and 
WMAP5 with $\Omega_m=0.258$ and $\sigma_8=0.796$.}, 
and introduce an updated theoretical model, which is integrated over 
the light-cone interval defined by the upper and lower redshift limits given 
by 2dFGRS data, and accounts for the scale-dependence of halo bias.


\begin{figure}
\includegraphics[width=84mm]{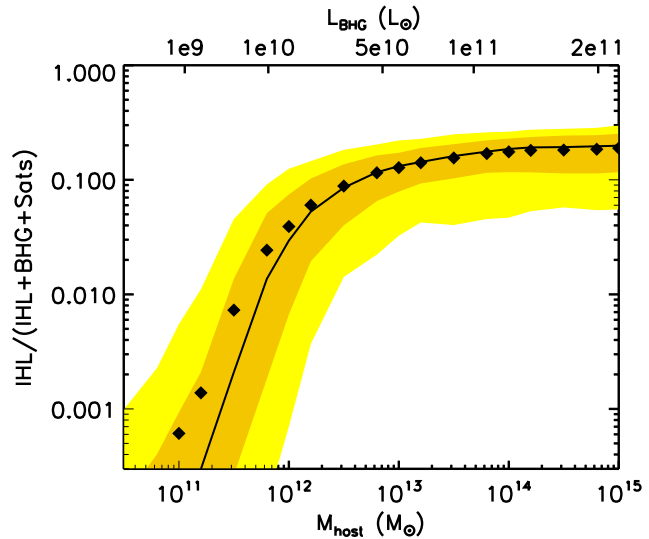}
\caption{
The predicted IHL fraction as a function of host halo virial 
mass ranging from $10^{10.5} \Msun \leq M_{host} \leq 10^{15} \Msun$.  The diamonds denote 
the mean of the distribution at fixed mass based on 1000 realizations 
of our analytic model, and the solid line marks the median.  The light shaded region 
encapsulates the 95\% range of the distribution at fixed mass, 
and the dark shaded region contains 68\% of the distribution.  The upper axes show the 
central galaxy (BHG) luminosity as derived from the $M_{host}/L_{BHG}$ mapping of \citet{vdb07}.
}
\label{fig:ihl2}
\end{figure} 

In Figure~\ref{fig:masslum}, we show the mass-to-light ratio derived by \citet{vdb07} and compare it 
to the previous CLF analysis of \citet{yang_etal03}, which was used in the fiducial model of PBZ07.  
In this paper, we update the model of \citet{Purcell_etal07} to reflect the more recent 
analysis of \citet{vdb07}.  The results of \citet{vdb07} indicate that in a WMAP1 cosmology 
galaxy formation occurs most efficiently in halos of host mass $M \sim 10^{11.5} \Msun$, where 
mass-to-light ratio is a minimum.  Otherwise, 
$M/L_c$ increases as a power-law with 
$M/L_c \propto M^{\gamma}$ where $\gamma=-2.32$ for 
systems with $M \la 10^{11.5} \hMsun$, and $\gamma=0.27$ for $M \ga 10^{11.5} \hMsun$.  
By comparison, the $M/L_c$ ratio derived by \citet{vdb07} for the WMAP3 
cosmology is 25-45\% lower across the spectrum of host mass, 
with the most pronounced difference occurring at 
the minimum near $M_{host} \sim 10^{11.5} \Msun$.  
We adopt the WMAP1 $M/L_c$ mapping in order to remain consistent with the cosmological 
parameters used in our previous work, but IHL fractions produced by our model in a WMAP3 cosmology 
differ by less than $1\%$ from the WMAP1 result on all mass scales, and we note that the 
intermediate WMAP5 parameters therefore also yield indistinguishable results for the IHL fraction.  
In addition, we draw our luminosity values according to Gaussian distributions of $M/L_c$ at fixed 
host mass, with a constant $\sigma = 0.4$ dex variance that roughly approximates 
the scatter in the probability distributions obtained by \citet{vdb07}, 
though our results are relatively robust to this choice.


\begin{figure}
\includegraphics[width=84mm]{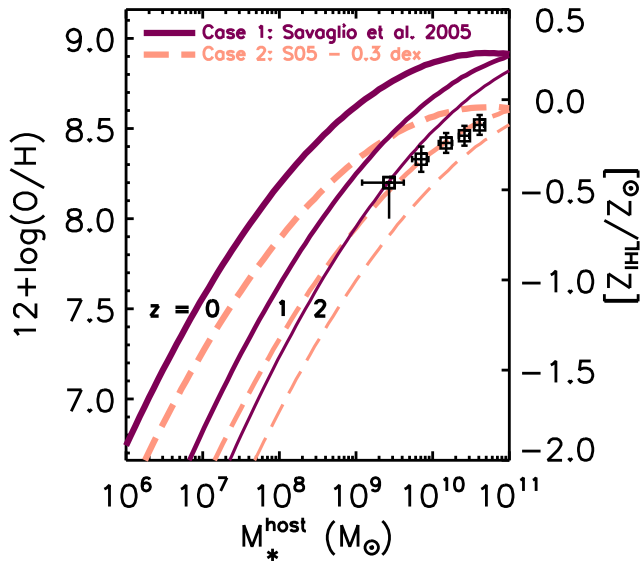}
\caption{
The fiducial, redshift-dependent mass-metallicity relation used to 
assign oxygen abundances to disrupted satellite galaxies produced by the model of PBZ07, 
in order to assess the chemical content of intrahalo luminosity.  
Drawn from the model of \citet{Savaglio_etal05} and formulated explicitly in Equation 1, 
the {\em maroon} lines represent case 1 of the 
$M_*-Z$ relation at $z=$0, 1, and 2 (solid lines of {\em high}, 
{\em medium}, and {\em low} thicknesses, respectively), 
and case 2 is plotted in {\em orange dashed} lines at the same redshifts, 
representing the original relation shifted by 0.3 dex towards 
lower metallicity, to reflect selection effects discussed in \S\ref{subsec:assigning}.  For 
comparison, we overplot with {\em black squares} data from the mass-metallicity 
relation at $z\sim 2$, as derived by \citet{Erb_etal06}. 
}
\label{fig:massmetal}
\end{figure} 

A less significant alteration to the PBZ07 model involves 
additional recursion levels applied to the EPS merger-tree algorithm, so that we now model the orbital 
evolution of subhalos-of-subhalos at each host mass, 
allowing us to probe diffuse light already present in an infalling system.  Numerical simulations 
suggest this pre-processed IHL may be a dominant contributor 
to the mild increase in the diffuse light fraction as a function of increasing host mass on cluster 
scales \citep{Murante_etal04,murante07,Monaco_etal06}, 
as observed by surveys testing the evolution of intracluster light with cluster richness \citep[\eg][]{Zibetti05}.  
In Figure~\ref{fig:ihl2}, we present updated results for the relative fraction of luminosity in 
diffuse IHL compared to the total luminosity in the system, which includes the contributions from the 
brightest halo galaxy (BHG) in the center as well as destroyed (IHL) {\em and} surviving satellites (Sats), as a 
function of host halo mass.  We predict a positive correlation between halo mass and IHL fraction 
spanning more than an order of magnitude in IHL fraction, with the dominant factor in the variation 
between this model and the result of PBZ07 being the updated slopes and minimum value in the behavior of 
$M/L_c(M_{host}$ as presented by \citet{vdb07}.  This prediction 
yields distributions of intrahalo light roughly consistent with 
galactic stellar halo luminosities, $\fihl \la 1-5\%$ for $M_{host} \la 10^{12} \Msun$, as well 
as intracluster observations of diffuse light, $\fihl \sim 10-40 \%$ for $M_{host} \sim 10^{14}-10^{15} \Msun$.


\begin{figure*}
\includegraphics{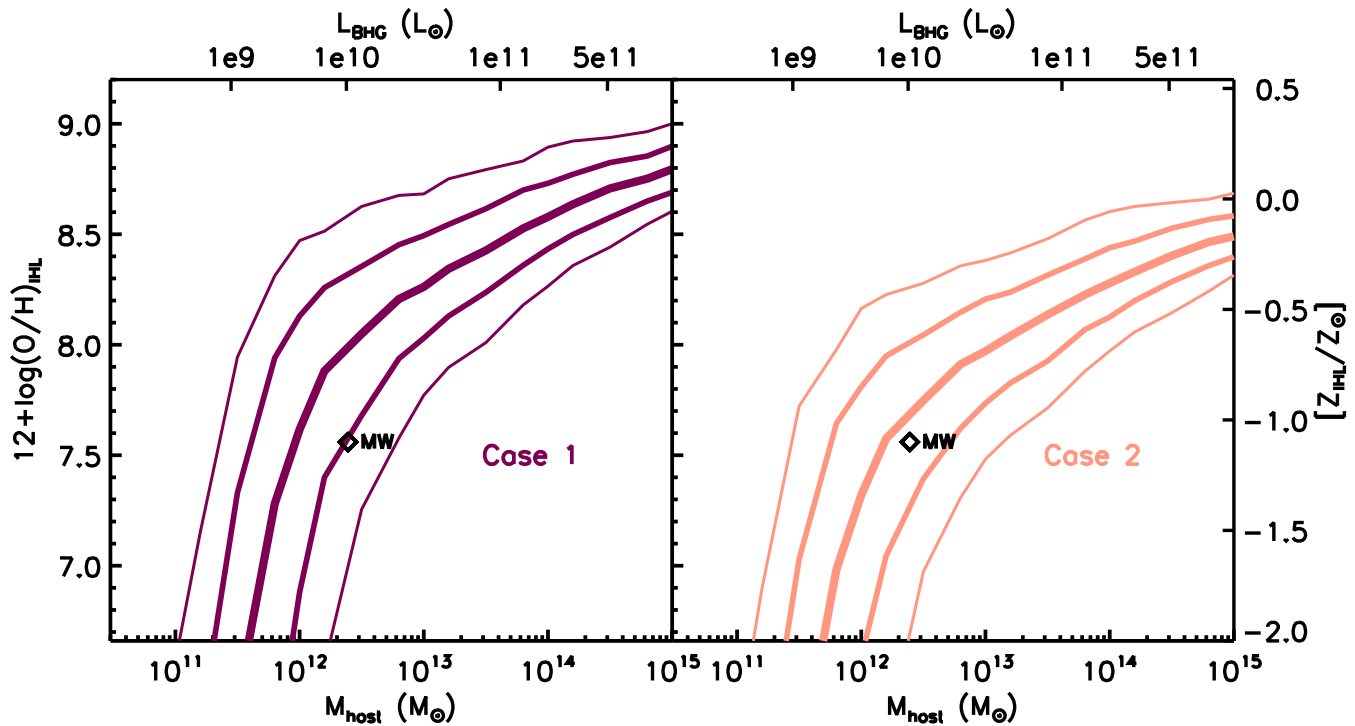}
\caption{
The oxygen abundance, relative to the solar value, 
for diffuse intrahalo stars in systems of mass $10^{10.5} \Msun \leq M_{host} \leq 10^{15} \Msun$, in 
both cases of the model.  The {\em doubly thick} 
solid line in each panel represents the mean of the 
$\ohihl$ distribution at each host mass.  The {\em thick solid} and 
{\em thin solid} lines denote the boundaries containing 
$68\%$ and $95\%$ of the distribution, respectively.  The observational estimate for the Milky Way 
stellar halo is shown with a {\em diamond}, $Z \sim -1.5$ as reported by \citet{Ferguson07}, with an 
added $\alpha$-enrichment of +0.4 \citep{mcwilliam97}.  The Galactic virial mass is taken to be
$2.43 \times 10^{12} \Msun$ as derived by \citet{Li_White08}.
}
\label{fig:result}
\end{figure*} 

\subsection{Assigning Metallicities}
\label{subsec:assigning} 

Having predicted the relative amount of
intrahalo stellar material in systems ranging  in mass from dwarf
galaxies up to rich galaxy clusters, we can  also determine the
expected chemical abundances in these diffuse components
by considering the well-known correlation between luminosity  and
metallicity (L-Z relation) in star-forming galaxies
\citep[\eg][]{garnettshields87,brodiehuchra91}.  The L-Z relation,
which holds over ten magnitudes in optical  luminosity
\citep{Zaritsky_etal94}, is generally thought to be the result of an
underlying correlation between stellar mass  and metallicity for
galaxies.  These two properties are fundamental relics of galaxy
formation and evolution, because they reflect the amount of gas tied up
in  a system's stars and the efficiency of gas recycling and retention
processes in that system, respectively.  

Resolving the age-metallicity-reddening  degeneracy for a particular stellar
population translates a galaxy's luminosity into stellar mass, $\Mstar$.  
The $\Mstar-Z$ correlation is ostensibly a consequence
of the interplay between a host halo's total baryonic mass and the
effective yield of the system; larger galaxies being less susceptible (having higher
stellar mass quotients) to outflow processes that decrease the quantities of metals
present for subsequent generations of star formation
\citep{Tremonti_etal04}.

Extensive surveys of star-forming galaxies have constrained the
mass-metallicity relation at various redshifts.  Using oxygen
gas-phase abundances derived from  optical nebular emission lines in galaxy data
drawn from the Sloan Digital Sky Survey (SDSS),
\citet{Tremonti_etal04} infer an $\Mstar-Z$ relationship for the local
universe ($z \la 0.1$), which is analytically well-fitted by a
second-order polynomial in log($\Mstar$).  This trend applies  for
galaxies with stellar masses between $10^8 \Msun \la \Mstar \la
10^{11.5}\Msun$, and the work of \citet{Lee_etal06} extends the
correlation into the  low-mass regime ($10^6 \Msun \la \Mstar \la
10^{9.5} \Msun$) by measuring 4.5 $\mu$m luminosities for nearby dwarf
irregular galaxies observed  by near-infrared instrumentation aboard
the Spitzer Space Telescope.  For the mass range in which the two
determinations overlap, the SDSS relation is roughly 0.3~dex  more
metal-rich than that derived for the dwarf-galaxy data, an offset that
may be partially explained by two selection effects: SDSS fiber
spectra preferentially  sample the innermost regions of spiral
galaxies which may in actuality have radial metallicity gradients
\citep[\eg][]{Zaritsky_etal94}, and oxygen abundances  derived for
SDSS data use bright-line estimation methods (in the absence of the
temperature-sensitive [O III] emission line) which overestimate the
true metallicity values  by $\sim$0.3 dex at roughly solar abundance
and greater \citep[\eg][]{kennicutt_etal03,bresolin_etal04}.


\begin{figure*}
\includegraphics{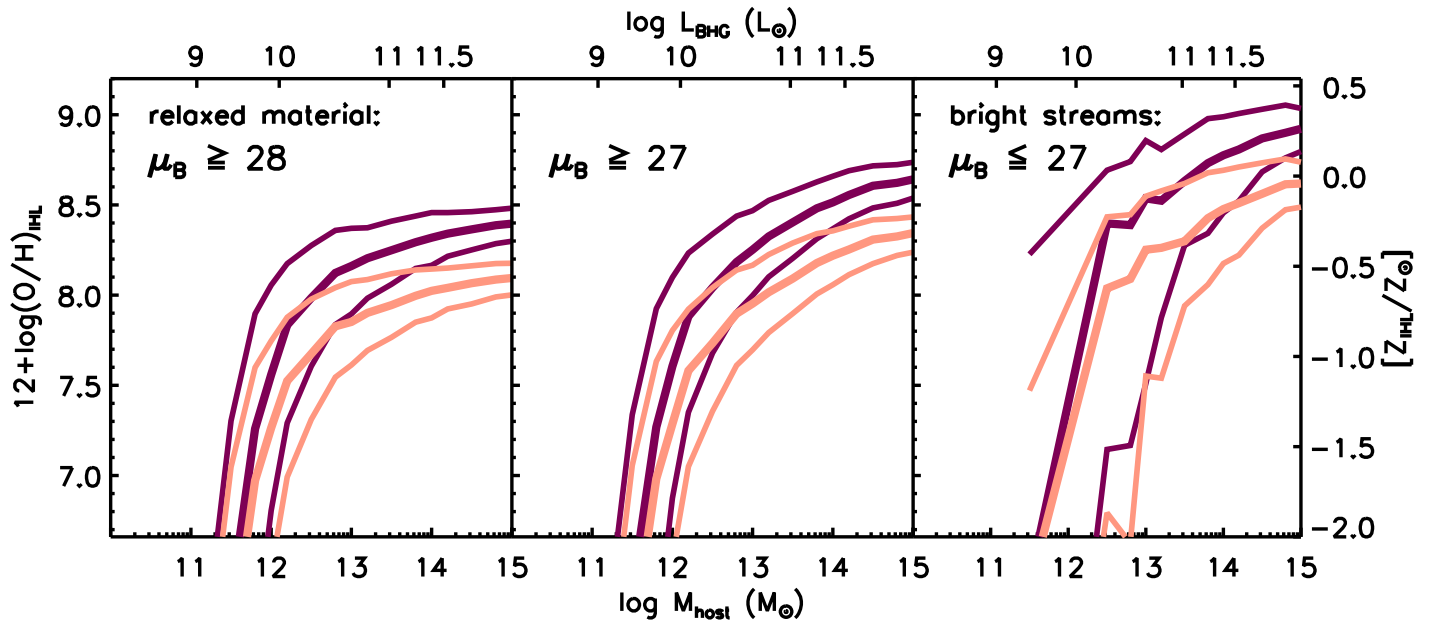}
\caption{ The $\ohihl$ distributions as a function of host mass for
diffuse stellar populations above or below a threshold in surface
brightness.  In the left panel,  we select dynamically older intrahalo
light by only showing the IHL metallicity for disrupted subhalos
having dimmed to $\mu_B \geq 28$ mag/arcsec$^2$, while  the center
panel includes diffuse material less bright than a slightly higher
threshold of $\mu_B \geq 27$ mag/arcsec$^2$.  By contrast, the right
panel selects only  dynamically young IHL features that remain
brighter than $\mu_B = 27$ mag/arcsec$^2$.  The {\em thick solid}
lines denote the mean of  the distribution at fixed host mass, while
the {\em thin solid} lines are contours bracketing $68\%$ of each
distribution.  }
\label{fig:brightcuts}
\end{figure*}

There is evidence that the mass-metallicity relation has evolved with
redshift.  Specifically, galaxies with a given stellar mass at
high redshift tend  to have lower oxygen abundances than their local
counterparts with the same stellar mass
\citep{Savaglio_etal05,Erb_etal06}.  Armed with a mass accretion
history for disrupted subhalos, we can assign their stellar mass
values using the adapted PBZ07 model and determine the appropriate
oxygen abundance of the resultant IHL component by applying an
$\Mstar-Z$ relation that evolves with redshift to match each satellite's
epoch of destruction.  To  accomplish this goal we adopt the
prescription of \citet[][hereafter S05]{Savaglio_etal05}, in which the
authors investigate stellar mass and metallicity in galaxies  at
redshift $0.4 \le z \le 1.0$ from the Gemini Deep Deep Survey (GDDS)
and Canada-France Redshift Survey (CFRS), in order to derive the form
by which the  $\Mstar-Z$ relation changes as a function of cosmic time.
This model translates the $\Mstar-Z$ polynomial of
\citet{Tremonti_etal04} into one consistent with  their choice of
stellar IMF and metallicity calibration, and extends the relation in
redshift.  It is worth noting here that we determine stellar metallicity 
via the time-dependent formulation describing oxygen gas-phase 
abundance, such that the chemical composition of our stars will reflect the 
metallicity of the gas from which they were formed, instead of the 
systematically richer gas-phase metallicity that would be measured at 
the present day.

To illustrate the importance of the selection-effect issues discussed above, 
we explore two mass-metallicity relations in our work.  We present 
predictions for IHL metallicity based on each of these mass-metallicity 
relations.  The two $\Mstar-Z$ relations represent relatively high- and 
low-metallicities respectively and we expect that the predictions 
from these two relations bracket the true value.  
In case 1, we adopt the model of S05, and in case 2
we decrease the oxygen abundance by 0.3 dex across all 
stellar masses as motivated by the offset found in the
analysis of \citet{Lee_etal06}.  For each destroyed satellite galaxy
in our analytic formalism, we draw from a Gaussian distribution of
metallicities centered on the mean $\Mstar-Z$ relation and with a
dispersion constant over host halo mass and equal to $\sigma = 0.2$ dex
as derived by S05 for their sample of galaxies at $z \sim 0.7$.  In
Figure~\ref{fig:massmetal}, we plot the fiducial mean mass-metallicity
relation at $z=0$, 1, and 2 for both cases.  Throughout this work we
adopt the standard units for oxygen  metallicity in galaxies, 
12+log(O/H) in which the solar abundance is 8.66 \citep{asplund04}.  The 
functional form of our case 1 as drawn from S05 is, for a given stellar mass $\Mstar$ 
and Hubble time $t_H$:
\begin{align*}
12+\mathrm{log(O/H)}=&-7.59+2.5315~\mathrm{log}~\Mstar-0.09649~\mathrm{log}^2\Mstar \\
&+5.1733~\mathrm{log}~t_H-0.3944~\mathrm{log}^2t_H \\
&-0.403~\mathrm{log}~t_H~\mathrm{log}~\Mstar ~~~~~~~~~~~~~~~~~~~~~\mathrm{(1)}
\end{align*}
For this prescription, in order to reflect the buildup of both stellar mass and 
the associated chemical composition of subhalo stars, we adopt the $t_H$ at 
which the satellite galaxy has formed half of the stellar mass it has at the 
time of accretion, \ie the appropriate $\Mstar(t_H) = 0.5\Mstar(t_{acc})$.  As 
in PBZ07, we have assumed for simplicity that star formation is truncated 
upon the subhalo's accretion onto the host halo, via processes of gas dynamics 
such as ram pressure stripping or strangulation.

Remnants of relatively recent accretion events will naturally be more
metal-enriched than the diffuse light from ancient merger remnants, 
which give rise to a more spatially-uniform, 
nearly-virialized component of diffuse luminosity that should be 
more metal-poor than the tidal streams that are the coherent 
remnants of recent accretion events \citep{Font_etal06}.  
Our formalism defines subhalo
disruption by mass loss alone and we do not trace the subsequent 
evolution of the liberated material in a self-consistent manner.  
Consequently, we must introduce 
additional modeling in order to distinguish bright tidal features
from older relics of galaxy disruption.  In their paper on the
interpretation of diffuse streams as the  result of satellite
disruption around galaxies in a~\LCDM universe, \citet{jsb01} derive
the form for the time evolution of a debris trail's surface
brightness, as a function of the  progenitor satellite's mass, the
host halo mass, and the galactocentric radius of the satellite's
orbit.  Their analysis is based on the destruction of satellites on 
circular orbits, but we  utilize this approximate form
in order to test the viability of separating the IHL metallicity 
along thresholds in dynamical age, which could be useful in studies of
Local Group  diffuse features such as the giant southern stream of
M31, as we discuss in \S\ref{sec:discussion}. 


\begin{figure}
\includegraphics[width=84mm]{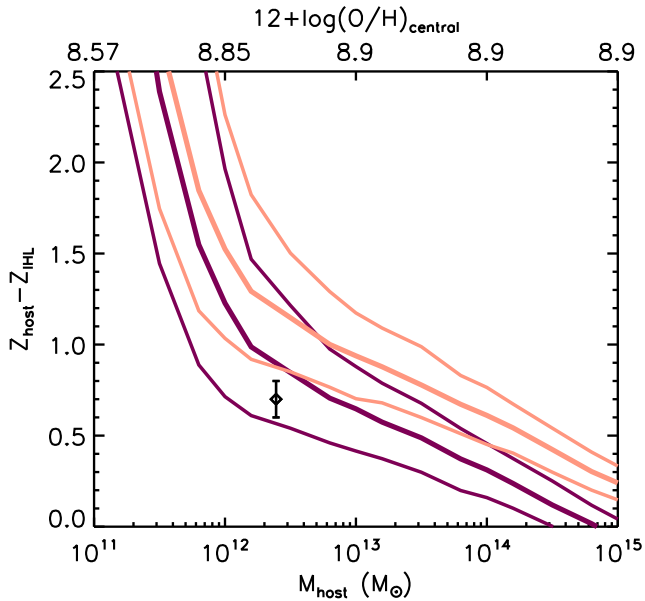}
\caption{
The difference between the metallicity of a host halo's central galaxy, $Z_{host}$, 
and the metallicity of the intrahalo component, $Z_{IHL}$.  For each case, the {\em thick solid}
lines denote the mean of the distribution at fixed host mass, while
the {\em thin solid} lines are contours bracketing $68\%$ of each
distribution.  Shown with a {\em diamond} is the observed difference for the Milky Way, 
assuming a Galactic metallicity of $[Fe/H] \sim -0.7 \pm 0.1$ solar, as determined by 
\citet{Ivezic_etal08}, with an $\alpha$-enrichment 
of $\sim +0.3$ as suggested by \citet{Chang_etal02} yielding $Z_{host} \sim -0.4 \pm 0.1$.  
The stellar halo metallicity is taken to be $Z_{IHL} \sim -1.1$ as discussed in 
\S~\ref{sec:discussion}.  The upper axis shows the central galaxy metallicities for each 
decade in host halo mass, as drawn from the fiducial $M_{host}-\Mstar$ relation and the 
case 1 mass-metallicity model.
}
\label{fig:zhostminuszihl}
\end{figure}

 \section{Results}
\label{sec:results}

We make predictions for the oxygen metallicities of diffuse intrahalo
stellar mass embedded in host halos with virial mass $10^{10.5} \Msun
\leq M_{host} \leq 10^{15} \Msun$, as shown explicitly in Figure~\ref{fig:result}.  
In case 1,  corresponding to the
redshift-dependent $\Mstar-Z$ relation obtained by S05, our analysis
predicts a mean $\ohihl$ of $\sim -0.2 - 0.2$ with respect to the solar
abundance, for diffuse intracluster luminosity present in halos with
mass larger than $M_{host} \sim 10^{14} \Msun$, while the case 2
metallicity is more sub-solar at those scales, $\ohihl \sim -0.5 - 0.0$  as
expected due to the difference between the normalizations of the two
fiducial models.  On galactic scales, we expect the abundance of the
stellar halo to rise quickly as a function of host mass,  increasing
from $\ohihl \sim -2$ to $\ohihl \sim -1$ in the range $10^{11.5}
\Msun \la M_{host} \la 10^{12} \Msun$.  IHL metallicities in both
cases evolve much more slowly in the regime of  intragroup and
intracluster light, the trend growing more shallow for host halos larger than
$M_{host} \sim 10^{12.5}$ and rising weakly from $\ohihl \sim -0.5$ to
roughly solar metallicities, inducing a  more distinct separation
between case 1 and case 2, whose $\ohihl$ distributions bracket
the solar abundance on the largest scales investigated.

The distribution of IHL metallicity at fixed host mass is quite wide
for stellar halos, with variance $\sigma_Z \sim 0.5$ dex at
Galactic-mass scales, and is significantly more narrow for intragroup
and intracluster light, where $\sigma_Z \sim 0.1$ dex and the average
chemical abundance of diffuse stellar mass reaches a plateau.  The
normalization of this plateau varies non-trivially as we  consider
imposing thresholds in surface brightness, corresponding to the
selection by dynamical age of destroyed satellite galaxies in order to
distinguish the old, metal-poor intrahalo component from  younger,
brighter and more metal-rich stellar substructure currently undergoing
disruption.  In the left panel of Figure~\ref{fig:brightcuts}, we plot
our case 1 IHL metallicity as a function of host mass  for only those
subhalos whose tidal streams have dimmed to a B-band surface
brightness lower than $\mu_B = 28$ mag/arcsec$^2$, and in the center
panel we show the same trend  for diffuse features dimmer than $\mu_B
= 27$ mag/arcsec$^2$.  By contrast, the right panel excludes faint
subhalo debris, selecting dynamically young streams that remain
brighter than  $\mu_B = 27$ mag/arcsec$^2$, and thus becoming noisy 
at the galactic mass scale, since not every realization in this regime has 
had such an event.  

\section{Discussion}
\label{sec:discussion}

One explanation for the origin of the mass-metallicity relation for
star-forming galaxies arises from galactic winds that induce a
significant amount of metal-enhanced mass loss from galaxies with
stellar mass below roughly $10^{10.5}$ \citep[][see also
\citealt{Kauffmann_etal04} regarding this threshold's importance for
galaxy evolution in general]{Tremonti_etal04,Gallazzi_etal05}, with
the "blowout" process becoming  more efficient as the system's host
mass decreases, although \citet{Lee_etal06} endorse a more leisurely
model for metal outflows due to the small variance they obtain in the
$\Mstar-Z$ relationship for dwarf  irregular galaxies.  Below the
critical scale $\Mstar^{crit} \sim 10^{10.5} \Msun$, the $\Mstar-Z$
correlation's slope grows steeper with lookback time, supporting a
galactic chemical evolution model predicated on longer e-folding
times for star formation in lower-mass galaxies
\citep{Savaglio_etal05}, a hypothesis consistent with the empirical
phenomenon of cosmic "downsizing" in which more massive galaxies form
the majority of their stars at a higher redshift than do less massive
systems \citep{Heavens_etal04,Juneau_etal05,PerezGonzalez_etal07}.


\begin{figure*}
\includegraphics{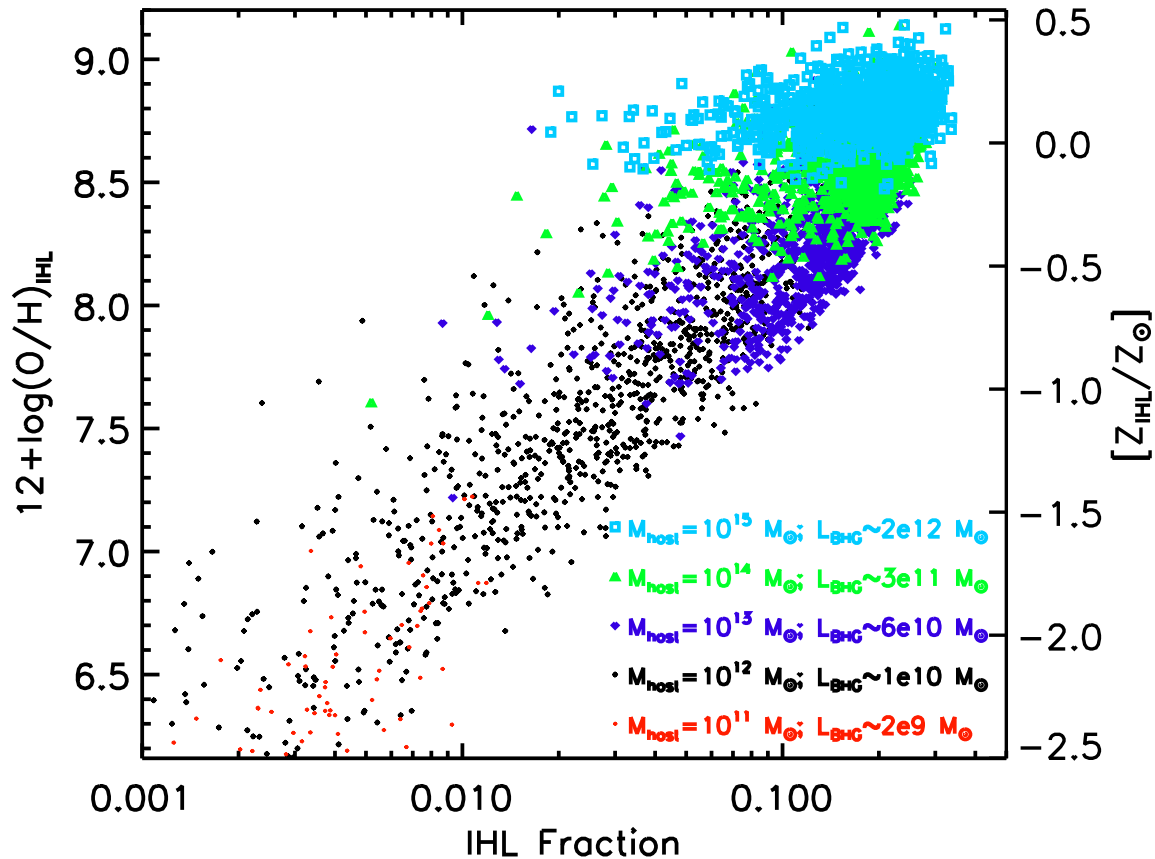}
\caption{
Case 1 $\ohihl$ as a function of IHL fraction for each 
realization at various fixed host masses.  
}
\label{fig:ohvsihl}
\end{figure*}

Numerical hydrodynamical simulations of hierarchical structure
formation have also generated a characteristic stellar mass
$\Mstar^{crit} \sim 3 \times 10^{10} \Msun$ which divides galaxy
populations into  two groups. The first contains large 
systems that formed most of their
stars more than $\sim 10$ Gyr ago, having been assembled via mergers
of stellar-dominated subhalos with little gas left over to produce a
significant  starburst during coalescence.  The second group is composed 
of comparatively small
systems that form stars more passively, allowing gas-rich mergers to
produce enough star formation that the metallicity of the overall
stellar content is affected by  the significant fraction of new stars
\citep[][see also \citealt{Tissera_etal05}]{DeRossi_etal07}.  This
stellar mass threshold remains constant to a redshift of $z \sim 3$,
and $M_*^{crit}$ galaxies have roughly solar  abundances that evolve
very weakly over the same timeframe, these systems being typically
embedded in halos of mass $M_{host} \sim 10^{12.5} \Msun$ according to
the formalism of \citet{vdb07}.  Convolving this expectation with our
prediction  regarding the metallicity of diffuse intrahalo light
produced by galaxy disruption, we note that both simulations
\citep{Stewart_etal08} and analytic models of galaxy formation
\citep{Purcell_etal07} describe \LCDM structure  on all scales as
being preferentially formed via mergers involving subhalos of mass $M
\sim 0.1 M_{host}$, indicating that both accreted substructure and the
diffuse intrahalo component should have roughly solar  metallicities
in halos of mass $M_{host} \sim 10^{13.5} \Msun$.  This context
slightly favors our case 1 result, in which the mean metallicity of
intragroup light on the relevant scale is very close to the solar
oxygen abundance, although there is a scatter $\sigma \sim 0.15$ dex
at fixed host mass in both cases.

Recent measurements of stellar halo metallicity in the Local Group
seem to cluster around the point Z$ \sim -1.5$ \citep[][see also
\citealt{Kalirai06} for a review of the halo abundance of
M31]{Ferguson07}.   However, the alpha-enrichment of the Galactic halo
is significant \citep[observed at +0.4 as reviewed by][]{mcwilliam97},
which is consistent with an intrahalo stellar population forming in an
early  burst, before type Ia supernovae could contribute much iron to
the IGM \citep[][see \citealt{venn_etal04} for a review of chemical
signatures in hierarchical formation models]{matteucci2003}.  An
appropriate observational benchmark for the stellar halo oxygen
abundance is therefore [O/H]$_{IHL} \sim -1.1$ with respect to the
solar metallicity, which falls along the lower 68\% contour in case 1 of our
fiducial theoretical result.  For the Galactic disk's iron abundance of $[Fe/H] 
\sim -0.7 \pm 0.1$ with respect to the solar metallicity \citep{Ivezic_etal08}, 
and an $\alpha$-enrichment of $\sim +0.3$ as suggested by \citet{Chang_etal02}, 
we also note that the difference between this $Z_{MW} \sim -0.4 \pm 0.1$ and 
the IHL metallicity is $\sim 0.7$ dex.  We show the expected behavior of the quantity $Z_{host}-Z_{IHL}$ 
as a function of host mass in Figure~\ref{fig:zhostminuszihl} and note that the 
Milky Way's observed value slightly favors our case 1 model; future 
abundance surveys of distant galaxies and diffuse light may provide further
empirical constraints supporting the prediction that intracluster stars have similar metallicities to 
that of the cluster's central galaxy, whereas intrahalo populations on smaller scales 
are significantly more metal-poor than their host galaxies.

The evolution of a galaxy's metallicity arises from the equilibrating processes 
of the infall of gas and its subsequent enrichment in metals via star formation; these 
competing rates of dilution and enhancement are governed by the efficiency of gas 
accretion and processing.  Recent theoretical investigations have reproduced observed metallicities in the 
intergalactic medium at $z \ga 2$ and matched the $\Mstar-Z$ relation's 
slope and normalization by introducing a paradigm in which galactic wind velocities 
are proportional to the stellar velocity dispersion of the host system \citep[the ``momentum-driven" 
scenario of][]{Finlator_Dave2008} and the mass-loading factor $\eta$ of the outflow scales 
with a galaxy's stellar mass as $\eta \propto \Mstar^{-1/3}$ at low masses and approaches 
unity at the blowout point, where winds begin to escape their host halo and the $\Mstar-Z$ 
relation flattens to a roughly constant value. 
The steep rise in IHL abundance at and below the galactic scale occurs as intragalactic
metal depletion processes dramatically fall to minimal efficiency with increasing host
halo mass, thereby enriching satellite systems prior to disruption,
and thus increasing the metallicity of the diffuse stellar component
in higher-order structure as a similarly fast function of host mass.
Given the rapidity of this trend, and the significant variance we
obtain in the metallicities of stellar halos belonging to spiral-sized
galaxies, we note that observational constraints on a possible
correlation between IHL metallicity and host galaxy luminosity will
likely reflect  the stochastic and ongoing nature of the diffuse
stellar component's formation, as well as the particular mass
accretion history of a given galaxy.  

One such relationship is advocated by \citet{Mouhcine_etal05}, in
which stellar halo abundances are derived from color measurements of
the red giant branch in off-disk fields and shown to increase with
galactic luminosity across  a wide range $-2 \la Z \la -0.5$, although
\citet{Ferguson07} claims this correlation is non-existent when halo
stars are kinematically selected.  A more recent application of this
method to an RGB-star survey of the Galaxy-analogue system NGC~891
has shown a broad IHL abundance distribution centered around $Z =
-0.9$, more metal-rich by a factor of three than the stellar halos of
both large Local Group galaxies \citep{Mouhcine_etal07}.  We do
predict a strong rise in halo metallicity as host mass increases
through the galactic scale (shown in Figure~\ref{fig:result}), but the
wide scatter at  fixed mass suggests that the statistical accuracy of
any such empirically-derived relation would be dramatically
contaminated by variance in the individual mass assembly processes for
each galaxy, as demonstrated by Figure~\ref{fig:ohvsihl} in which we
obtain a power-law relationship between IHL abundance and the fraction
of a system's stellar mass contributed by the intrahalo component ($n
\sim 0.95$ in $\ohihl \propto n$~log~$\fihl$), spanning the widest
range at galactic scales and thereby suggesting that stellar halo
metallicity more properly correlates with the luminosity of the
stellar halo, and not necessarily the luminosity of the parent
galaxy. 

This correlation grows weaker when we consider increasingly more
massive host halos, as the IHL metallicity distribution tightens and
the IHL fraction flattens.  We predict an $\ohihl$ value of $\sim -0.5$ 
for weak groups such as the Local Group, which has a mass $M_{host} 
\sim 5.3 \times 10^{12} \Msun$ according to \citet{Li_White08}, 
although observational constraints in this regime remain elusive.  
On still larger scales, the Virgo cluster
provides a more useful laboratory for the study of intrahalo stellar
metallicity, although systematic errors in the photometry of RGB-tip
stars in these fields cause great variance in the abundance values
yielded \citep[\eg][in which the authors obtain $-0.8 \la Z \la
-0.2$]{Durrell_etal02}.  In a recent study, \citet{Williams_etal07}
identify $\sim5300$ intracluster stars in Virgo, determining that
$70\%-80\%$ of them are older than $\sim10$ Gyr and populated over a
wide range in abundance ($-2.5 \la Z \la 0.0$), as we might expect
from a diffuse population with multiple origins; the remaining
$20\%-30\%$ of the intrahalo population has a metallicity distribution
function peaking near solar values, these younger stars having been
more recently orphaned by disruption processes acting on their parent
galaxy.  Still more metal-rich is the intrahalo population of the
cluster Abell 3888, which contains one component having formed at
redshift $z > 1$ with high metallicity ($1 < Z \la 2.5$), as well as a
more centralized diffuse region at roughly solar abundance
\citep{Krick_etal06}.  The variation in our predicted distribution is
of order $\sigma \sim 0.1$ dex, although the location of the plateau
varies significantly with surface brightness cuts intended to separate
older intrahalo populations from dynamically young IHL plumes in
galaxy clusters.  Typical measurements of intracluster light have
analyzed fields without obvious luminous substructure, corresponding
to the spheroidal halo of field stars in spiral galaxies. It appears
that we must be careful about claims as to how much stellar mass
actually lies in long-disrupted satellite remnants, as opposed to
relatively bright diffuse tidal streams. 

However, it is worth noting that stellar halo observations of galaxies
outside the Local Group cannot distinguish 
stream-like substructure from background field stars, just as
intragroup and intracluster studies cannot.  Tidal features such as
M31's giant southern stream are rich in stellar mass \citep[the
progenitor may have been as bright as a billion solar luminosities
according to][]{font_etal06m31}  as well as metals 
\citep[$Z \sim -0.5$ according to][]{Guhathakurta_etal05}, so we can
reconcile our predictions with current galactic measurements given the
caveat that typical halo metallicities reflect  the underlying stellar
component, without the inclusion of bright outer-halo substructure
resulting from massive accretion events, a distinction made more clear
by a recent semi-analytic treatment of baryon  dynamics in numerical
models of L$_*$-galaxy halos \citep[][see also
\citealt{BullockJohnston05} for details of the
simulation]{Font_etal07}.  We also predict a clear correlation between
the oxygen abundance and the surface brightness of tidal streams in
galaxy clusters, a prediction that will be testable as future deep
surveys come online and metallicity distribution functions become more
well-studied in diffuse intrahalo stellar populations at and above the
galaxy scale.

$\;$

We would like to thank Betsy Barton and Brant Robertson for useful discussions.  
CWP and JSB are supported by National Science Foundation (NSF) 
grants AST-0607377 and AST-0507816, and the Center for 
Cosmology at UC Irvine.  ARZ is funded by the University of Pittsburgh.


\begin{thebibliography}{}


\bibitem[{{Abadi} {et~al.}(2006){Abadi}, {Navarro}, \&
  {Steinmetz}}]{Abadi_etal06}
{Abadi}, M.~G., {Navarro}, J.~F., \& {Steinmetz}, M. 2006, MNRAS, 365, 747

\bibitem[{{Asplund} {et~al.}(2004){Asplund}, {Grevesse}, {Sauval}, {Allende
  Prieto}, \& {Kiselman}}]{asplund04}
{Asplund}, M., {Grevesse}, N., {Sauval}, A.~J., {Allende Prieto}, C., \&
  {Kiselman}, D. 2004, A\&A, 417, 751

\bibitem[{{Bond} {et~al.}(1991){Bond}, {Cole}, {Efstathiou}, \&
  {Kaiser}}]{Bond91}
{Bond}, J.~R., {Cole}, S., {Efstathiou}, G., \& {Kaiser}, N. 1991, ApJ, 379,
  440

\bibitem[{{Bresolin} {et~al.}(2004){Bresolin}, {Garnett}, \&
  {Kennicutt}}]{bresolin_etal04}
{Bresolin}, F., {Garnett}, D.~R., \& {Kennicutt}, Jr., R.~C. 2004, ApJ, 615,
  228

\bibitem[{{Brodie} \& {Huchra}(1991)}]{brodiehuchra91}
{Brodie}, J.~P. \& {Huchra}, J.~P. 1991, ApJ, 379, 157

\bibitem[{{Bullock} \& {Johnston}(2005)}]{bj05}
{Bullock}, J.~S. \& {Johnston}, K.~V. 2005, ApJ, 635, 931

\bibitem[{{Bullock} \& {Johnston}(2005)}]{BullockJohnston05}
---. 2005, ApJ, 635, 931

\bibitem[{{Bullock} {et~al.}(2001){Bullock}, {Kravtsov}, \&
  {Weinberg}}]{Bullock_etal01}
{Bullock}, J.~S., {Kravtsov}, A.~V., \& {Weinberg}, D.~H. 2001, ApJ, 548, 33

\bibitem[{{Byrd} \& {Valtonen}(1990)}]{Byrd_Valtonen90}
{Byrd}, G. \& {Valtonen}, M. 1990, ApJ, 350, 89

\bibitem[{{Calc{\'a}neo-Rold{\'a}n} {et~al.}(2000){Calc{\'a}neo-Rold{\'a}n},
  {Moore}, {Bland-Hawthorn}, {Malin}, \& {Sadler}}]{cmbs00}
{Calc{\'a}neo-Rold{\'a}n}, C., {Moore}, B., {Bland-Hawthorn}, J., {Malin}, D.,
  \& {Sadler}, E.~M. 2000, MNRAS, 314, 324
  
\bibitem[Chang et al.(2002)]{Chang_etal02} Chang, R.-X., Shu, C.-G., 
\& Hou, J.-L.\ 2002, Chinese Journal of Astronomy and Astrophysics, 2, 226 

\bibitem[{{Chapman} {et~al.}(2006){Chapman}, {Ibata}, {Lewis}, {Ferguson},
  {Irwin}, {McConnachie}, \& {Tanvir}}]{Chapman_etal06}
{Chapman}, S.~C., {Ibata}, R., {Lewis}, G.~F., {Ferguson}, A.~M.~N., {Irwin},
  M., {McConnachie}, A., \& {Tanvir}, N. 2006, astro-ph/0602604

\bibitem[{{Chiba} \& {Beers}(2000)}]{Chiba_Beers00}
{Chiba}, M. \& {Beers}, T.~C. 2000, AJ, 119, 2843

\bibitem[Conroy et al.(2007)]{Conroy_etal07} Conroy, C., Wechsler, 
R.~H., \& Kravtsov, A.~V.\ 2007, ApJ, 668, 826 

\bibitem[{{Cooray} \& {Milosavljevi{\'c}}(2005)}]{cm:05}
{Cooray}, A. \& {Milosavljevi{\'c}}, M. 2005, ApJL, 627, L85

\bibitem[{{De Lucia} \& {Helmi}(2008)}]{DeLuciaHelmi08} De Lucia, G., \& Helmi, 
A.\ 2008, ArXiv e-prints, 804, arXiv:0804.2465 

\bibitem[{{de Rossi} {et~al.}(2007){de Rossi}, {Tissera}, \&
  {Scannapieco}}]{DeRossi_etal07}
{de Rossi}, M.~E., {Tissera}, P.~B., \& {Scannapieco}, C. 2007, MNRAS, 374,
  323

\bibitem[{{Diemand} {et~al.}(2005){Diemand}, {Madau}, \&
  {Moore}}]{Diemand_etal05}
{Diemand}, J., {Madau}, P., \& {Moore}, B. 2005, MNRAS, 364, 367

\bibitem[{{Dubinski} {et~al.}(2003){Dubinski}, {Koranyi}, \&
  {Geller}}]{Dubinski03}
{Dubinski}, J., {Koranyi}, D., \& {Geller}, M. 2003, in IAU Symposium, ed.
  J.~{Makino} \& P.~{Hut}, 208--237

\bibitem[{{Durrell} {et~al.}(2002){Durrell}, {Ciardullo}, {Feldmeier},
  {Jacoby}, \& {Sigurdsson}}]{Durrell_etal02}
{Durrell}, P.~R., {Ciardullo}, R., {Feldmeier}, J.~J., {Jacoby}, G.~H., \&
  {Sigurdsson}, S. 2002, ApJ, 570, 119

\bibitem[{{Erb} {et~al.}(2006){Erb}, {Shapley}, {Pettini}, {Steidel}, {Reddy},
  \& {Adelberger}}]{Erb_etal06}
{Erb}, D.~K., {Shapley}, A.~E., {Pettini}, M., {Steidel}, C.~C., {Reddy},
  N.~A., \& {Adelberger}, K.~L. 2006, ApJ, 644, 813

\bibitem[{{Feldmeier} {et~al.}(2004){Feldmeier}, {Ciardullo}, {Jacoby}, \&
  {Durrell}}]{Feldmeier_etal04}
{Feldmeier}, J.~J., {Ciardullo}, R., {Jacoby}, G.~H., \& {Durrell}, P.~R. 2004,
  ApJ, 615, 196

\bibitem[{{Ferguson}(2007)}]{Ferguson07}
{Ferguson}, A. 2007, astro-ph/0702224

\bibitem[Finlator 
\& Dav{\'e}(2008)]{Finlator_Dave2008} Finlator, K., \& Dav{\'e}, R.\ 2008, MNRAS, 385, 2181

\bibitem[{{Font} {et~al.}(2006){Font}, {Johnston}, {Bullock}, \&
  {Robertson}}]{Font_etal06}
{Font}, A.~S., {Johnston}, K.~V., {Bullock}, J.~S., \& {Robertson}, B.~E.
  2006, ApJ, 638, 585

\bibitem[{{Font} {et~al.}(2007){Font}, {Johnston}, {Ferguson}, {Bullock},
  {Robertson}, {Tumlinson}, \& {Guhathakurta}}]{Font_etal07}
{Font}, A.~S., {Johnston}, K.~V., {Ferguson}, A.~M.~N., {Bullock}, J.~S.,
  {Robertson}, B.~E., {Tumlinson}, J., \& {Guhathakurta}, P. 2007,
  arXiv:astro-ph/0709.2076

\bibitem[{{Font} {et~al.}(2006){Font}, {Johnston},
  {Guhathakurta}, {Majewski}, \& {Rich}}]{font_etal06m31}
{Font}, A.~S., {Johnston}, K.~V., {Guhathakurta}, P., {Majewski}, S.~R., \&
  {Rich}, R.~M. 2006, AJ, 131, 1436

\bibitem[{{Gallagher} \& {Ostriker}(1972)}]{Gallagher_Ostriker72}
{Gallagher}, III, J.~S. \& {Ostriker}, J.~P. 1972, AJ, 77, 288

\bibitem[{{Gallazzi} {et~al.}(2005){Gallazzi}, {Charlot}, {Brinchmann},
  {White}, \& {Tremonti}}]{Gallazzi_etal05}
{Gallazzi}, A., {Charlot}, S., {Brinchmann}, J., {White}, S.~D.~M., \&
  {Tremonti}, C.~A. 2005, MNRAS, 362, 41

\bibitem[{{Garnett} \& {Shields}(1987)}]{garnettshields87}
{Garnett}, D.~R. \& {Shields}, G.~A. 1987, ApJ, 317, 82

\bibitem[{{Gnedin}(2003)}]{Gnedin03}
{Gnedin}, O.~Y. 2003, ApJ, 582, 141

\bibitem[{{Gonzalez} {et~al.}(2005){Gonzalez}, {Zabludoff}, \&
  {Zaritsky}}]{Gonzalez_etal05}
{Gonzalez}, A.~H., {Zabludoff}, A.~I., \& {Zaritsky}, D. 2005, ApJ, 618, 195

\bibitem[{{Guhathakurta} {et~al.}(2005){Guhathakurta}, {Ostheimer}, {Gilbert},
  {Rich}, {Majewski}, {Kalirai}, {Reitzel}, \&
  {Patterson}}]{Guhathakurta_etal05}
{Guhathakurta}, P., {Ostheimer}, J.~C., {Gilbert}, K.~M., {Rich}, R.~M.,
  {Majewski}, S.~R., {Kalirai}, J.~S., {Reitzel}, D.~B., \& {Patterson}, R.~J.
  2005, astro-ph/0502366

\bibitem[{{Heavens} {et~al.}(2004){Heavens}, {Panter}, {Jimenez}, \&
  {Dunlop}}]{Heavens_etal04}
{Heavens}, A., {Panter}, B., {Jimenez}, R., \& {Dunlop}, J. 2004, Nature, 428, 625

\bibitem[{{Helmi} {et~al.}(1999)}]{Helmi_etal99} {Helmi}, A., {White}, 
S.~D.~M., {de Zeeuw}, P.~T., \& {Zhao}, H. 1999, Nature, 402, 53 
  625

\bibitem[{{Irwin} {et~al.}(2005){Irwin}, {Ferguson}, {Ibata}, {Lewis}, \&
  {Tanvir}}]{Irwin05}
{Irwin}, M.~J., {Ferguson}, A.~M.~N., {Ibata}, R.~A., {Lewis}, G.~F., \&
  {Tanvir}, N.~R. 2005, ApJL, 628, L105

\bibitem[{{Ivezi{\'c}} {et~al.}(2000){Ivezi{\'c}}, {Goldston}, {Finlator},
  {Knapp}, {Yanny}, {McKay}, {Amrose}, {Krisciunas}, {Willman}, {Anderson},
  {Schaber}, {Erb}, {Logan}, {Stubbs}, {Chen}, {Neilsen}, {Uomoto}, {Pier},
  {Fan}, {Gunn}, {Lupton}, {Rockosi}, {Schlegel}, {Strauss}, {Annis},
  {Brinkmann}, {Csabai}, {Doi}, {Fukugita}, {Hennessy}, {Hindsley}, {Margon},
  {Munn}, {Newberg}, {Schneider}, {Smith}, {Szokoly}, {Thakar}, {Vogeley},
  {Waddell}, {Yasuda}, \& {York}}]{Ivezic_etal00}
{Ivezi{\'c}}, {\v Z}., {Goldston}, J., {Finlator}, K., {Knapp}, G.~R., {Yanny},
  B., {McKay}, T.~A., {Amrose}, S., {Krisciunas}, K., {Willman}, B.,
  {Anderson}, S., {Schaber}, C., {Erb}, D., {Logan}, C., {Stubbs}, C., {Chen},
  B., {Neilsen}, E., {Uomoto}, A., {Pier}, J.~R., {Fan}, X., {Gunn}, J.~E.,
  {Lupton}, R.~H., {Rockosi}, C.~M., {Schlegel}, D., {Strauss}, M.~A., {Annis},
  J., {Brinkmann}, J., {Csabai}, I., {Doi}, M., {Fukugita}, M., {Hennessy},
  G.~S., {Hindsley}, R.~B., {Margon}, B., {Munn}, J.~A., {Newberg}, H.~J.,
  {Schneider}, D.~P., {Smith}, J.~A., {Szokoly}, G.~P., {Thakar}, A.~R.,
  {Vogeley}, M.~S., {Waddell}, P., {Yasuda}, N., \& {York}, D.~G. 2000, AJ,
  120, 963
  
  \bibitem[Ivezic et al.(2008)]{Ivezic_etal08} Ivezic, Z., et al.\ 
2008, ArXiv e-prints, 804, arXiv:0804.3850

\bibitem[{{Johnston}(1998)}]{Johnston_etal98}
{Johnston}, K.~V. 1998, ApJ, 495, 297

\bibitem[{{Johnston} {et~al.}(1996){Johnston}, {Hernquist}, \&
  {Bolte}}]{Johnston_etal96}
{Johnston}, K.~V., {Hernquist}, L., \& {Bolte}, M. 1996, ApJ, 465, 278

\bibitem[{{Johnston} {et~al.}(2001){Johnston}, {Sackett}, \& {Bullock}}]{jsb01}
{Johnston}, K.~V., {Sackett}, P.~D., \& {Bullock}, J.~S. 2001, ApJ, 557, 137

\bibitem[{{Juneau} {et~al.}(2005){Juneau}, {Glazebrook}, {Crampton},
  {McCarthy}, {Savaglio}, {Abraham}, {Carlberg}, {Chen}, {Le Borgne}, {Marzke},
  {Roth}, {J{\o}rgensen}, {Hook}, \& {Murowinski}}]{Juneau_etal05}
{Juneau}, S., {Glazebrook}, K., {Crampton}, D., {McCarthy}, P.~J., {Savaglio},
  S., {Abraham}, R., {Carlberg}, R.~G., {Chen}, H.-W., {Le Borgne}, D.,
  {Marzke}, R.~O., {Roth}, K., {J{\o}rgensen}, I., {Hook}, I., \& {Murowinski},
  R. 2005, ApJL, 619, L135

\bibitem[{{Kalirai} {et~al.}(2006){Kalirai}, {Gilbert}, {Guhathakurta},
  {Majewski}, {Ostheimer}, {Rich}, {Cooper}, {Reitzel}, \&
  {Patterson}}]{Kalirai06}
{Kalirai}, J.~S., {Gilbert}, K.~M., {Guhathakurta}, P., {Majewski}, S.~R.,
  {Ostheimer}, J.~C., {Rich}, R.~M., {Cooper}, M.~C., {Reitzel}, D.~B., \&
  {Patterson}, R.~J. 2006, astro-ph/0605170

\bibitem[{{Kauffmann} {et~al.}(1993){Kauffmann}, {White}, \&
  {Guiderdoni}}]{kwg93}
{Kauffmann}, G., {White}, S.~D.~M., \& {Guiderdoni}, B. 1993, MNRAS, 264, 201

\bibitem[{{Kauffmann} {et~al.}(2004){Kauffmann}, {White}, {Heckman},
  {M{\'e}nard}, {Brinchmann}, {Charlot}, {Tremonti}, \&
  {Brinkmann}}]{Kauffmann_etal04}
{Kauffmann}, G., {White}, S.~D.~M., {Heckman}, T.~M., {M{\'e}nard}, B.,
  {Brinchmann}, J., {Charlot}, S., {Tremonti}, C., \& {Brinkmann}, J. 2004,
  MNRAS, 353, 713

\bibitem[{{Kennicutt} {et~al.}(2003){Kennicutt}, {Bresolin}, \&
  {Garnett}}]{kennicutt_etal03}
{Kennicutt}, Jr., R.~C., {Bresolin}, F., \& {Garnett}, D.~R. 2003, ApJ, 591,
  801

\bibitem[{{Krick} {et~al.}(2006){Krick}, {Bernstein}, \&
  {Pimbblet}}]{Krick_etal06}
{Krick}, J.~E., {Bernstein}, R.~A., \& {Pimbblet}, K.~A. 2006, AJ, 131, 168

\bibitem[{{Lacey} \& {Cole}(1993)}]{LC93}
{Lacey}, C. \& {Cole}, S. 1993, MNRAS, 262, 627

\bibitem[{{Lee} {et~al.}(2006){Lee}, {Skillman}, {Cannon}, {Jackson}, {Gehrz},
  {Polomski}, \& {Woodward}}]{Lee_etal06}
{Lee}, H., {Skillman}, E.~D., {Cannon}, J.~M., {Jackson}, D.~C., {Gehrz},
  R.~D., {Polomski}, E.~F., \& {Woodward}, C.~E. 2006, ApJ, 647, 970
  
\bibitem[{{Li} \& {White}(2008)}]{Li_White08} {Li}, Y.-S., \& {White}, S.~D.~M. 2008, MNRAS, 384, 1459

\bibitem[{{Lin} \& {Mohr}(2004)}]{Lin_Mohr04}
{Lin}, Y.-T. \& {Mohr}, J.~J. 2004, ApJ, 617, 879

\bibitem[{{Matteucci}(2003)}]{matteucci2003}
{Matteucci}, F. 2003, Elemental Abundances in Old Stars and Damped
  Lyman-{$\alpha$} Systems, 25th meeting of the IAU, Joint Discussion 15, 22
  July 2003, Sydney, Australia, 15

\bibitem[{{McConnachie} {et~al.}(2006){McConnachie}, {Chapman}, {Ibata},
  {Ferguson}, {Irwin}, {Lewis}, {Tanvir}, \& {Martin}}]{McConnachie_etal06}
{McConnachie}, A.~W., {Chapman}, S.~C., {Ibata}, R.~A., {Ferguson}, A.~M.~N.,
  {Irwin}, M.~J., {Lewis}, G.~F., {Tanvir}, N.~R., \& {Martin}, N. 2006, ApJL,
  647, L25

\bibitem[{{McWilliam}(1997)}]{mcwilliam97}
{McWilliam}, A. 1997, ARA\&A, 35, 503

\bibitem[{{Merritt}(1983)}]{Merritt83}
{Merritt}, D. 1983, ApJ, 264, 24

\bibitem[{{Mihos}(2004)}]{Mihos04}
{Mihos}, J.~C. 2004, in Clusters of Galaxies: Probes of Cosmological Structure
  and Galaxy Evolution, ed. J.~S. {Mulchaey}, A.~{Dressler}, \& A.~{Oemler},
  277

\bibitem[{{Mihos} {et~al.}(2005){Mihos}, {Harding}, {Feldmeier}, \&
  {Morrison}}]{Mihos_etal05}
{Mihos}, J.~C., {Harding}, P., {Feldmeier}, J., \& {Morrison}, H. 2005, ApJL,
  631, L41

\bibitem[{Monaco {et~al.}(2006)Monaco, Murante, Borgani, \&
  Fontanot}]{Monaco_etal06}
Monaco, P., Murante, G., Borgani, S., \& Fontanot, F. 2006, The Astrophysical
  Journal, 652, L89

\bibitem[{{Morrison} {et~al.}(2000){Morrison}, {Mateo}, {Olszewski}, {Harding},
  {Dohm-Palmer}, {Freeman}, {Norris}, \& {Morita}}]{Morrison_etal00}
{Morrison}, H.~L., {Mateo}, M., {Olszewski}, E.~W., {Harding}, P.,
  {Dohm-Palmer}, R.~C., {Freeman}, K.~C., {Norris}, J.~E., \& {Morita}, M.
  2000, AJ, 119, 2254

\bibitem[{{Mouhcine} {et~al.}(2005){Mouhcine}, {Ferguson}, {Rich}, {Brown}, \&
  {Smith}}]{Mouhcine_etal05}
{Mouhcine}, M., {Ferguson}, H.~C., {Rich}, R.~M., {Brown}, T.~M., \& {Smith},
  T.~E. 2005, ApJ, 633, 821

\bibitem[{{Mouhcine} {et~al.}(2007){Mouhcine}, {Rejkuba}, \&
  {Ibata}}]{Mouhcine_etal07}
{Mouhcine}, M., {Rejkuba}, M., \& {Ibata}, R. 2007, MNRAS, 842

\bibitem[{{Murante} {et~al.}(2004){Murante}, {Arnaboldi}, {Gerhard}, {Borgani},
  {Cheng}, {Diaferio}, {Dolag}, {Moscardini}, {Tormen}, {Tornatore}, \&
  {Tozzi}}]{Murante_etal04}
{Murante}, G., {Arnaboldi}, M., {Gerhard}, O., {Borgani}, S., {Cheng}, L.~M.,
  {Diaferio}, A., {Dolag}, K., {Moscardini}, L., {Tormen}, G., {Tornatore}, L.,
  \& {Tozzi}, P. 2004, ApJL, 607, L83

\bibitem[{{Murante} {et~al.}(2007){Murante}, {Giovalli}, {Gerhard},
  {Arnaboldi}, {Borgani}, \& {Dolag}}]{murante07}
{Murante}, G., {Giovalli}, M., {Gerhard}, O., {Arnaboldi}, M., {Borgani}, S.,
  \& {Dolag}, K. 2007, astro-ph/0701925

\bibitem[{{Perez-Gonzalez} {et~al.}(2007){Perez-Gonzalez}, {Rieke}, {Villar},
  {Barro}, {Blaylock}, {Egami}, {Gallego}, {Gil de Paz}, {Pascual}, {Zamorano},
  \& {Donley}}]{PerezGonzalez_etal07}
{Perez-Gonzalez}, P.~G., {Rieke}, G.~H., {Villar}, V., {Barro}, G., {Blaylock},
  M., {Egami}, E., {Gallego}, J., {Gil de Paz}, A., {Pascual}, S., {Zamorano},
  J., \& {Donley}, J.~L. 2007, ArXiv e-prints, 709

\bibitem[{{Purcell} {et~al.}(2007){Purcell}, {Bullock}, \&
  {Zentner}}]{Purcell_etal07}
{Purcell}, C.~W., {Bullock}, J.~S., \& {Zentner}, A.~R. 2007, ApJ, 666, 20

\bibitem[{{Read} {et~al.}(2006){Read}, {Pontzen}, \& {Viel}}]{Read_etal06}
{Read}, J.~I., {Pontzen}, A.~P., \& {Viel}, M. 2006, MNRAS, 821

\bibitem[{{Robertson} {et~al.}(2005){Robertson}, {Bullock}, {Font}, {Johnston},
  \& {Hernquist}}]{Robertson_etal05}
{Robertson}, B., {Bullock}, J.~S., {Font}, A.~S., {Johnston}, K.~V., \&
  {Hernquist}, L. 2005, ApJ, 632, 872

\bibitem[{{Rudick} {et~al.}(2006){Rudick}, {Mihos}, \&
  {McBride}}]{Rudick_etal06}
{Rudick}, C.~S., {Mihos}, J.~C., \& {McBride}, C. 2006, astro-ph/0605603

\bibitem[{{Ryan} \& {Norris}(1991)}]{ryan_norris91}
{Ryan}, S.~G. \& {Norris}, J.~E. 1991, AJ, 101, 1865

\bibitem[{{Savaglio} {et~al.}(2005){Savaglio}, {Glazebrook}, {Le Borgne},
  {Juneau}, {Abraham}, {Chen}, {Crampton}, {McCarthy}, {Carlberg}, {Marzke},
  {Roth}, {J{\o}rgensen}, \& {Murowinski}}]{Savaglio_etal05}
{Savaglio}, S., {Glazebrook}, K., {Le Borgne}, D., {Juneau}, S., {Abraham},
  R.~G., {Chen}, H.-W., {Crampton}, D., {McCarthy}, P.~J., {Carlberg}, R.~G.,
  {Marzke}, R.~O., {Roth}, K., {J{\o}rgensen}, I., \& {Murowinski}, R. 2005,
  ApJ, 635, 260

\bibitem[{{Searle} \& {Zinn}(1978)}]{Searle_Zinn78}
{Searle}, L. \& {Zinn}, R. 1978, ApJ, 225, 357

\bibitem[{{Seigar} {et~al.}(2006){Seigar}, {Barth}, \&
  {Bullock}}]{Seigar_etal06}
{Seigar}, M.~S., {Barth}, A.~J., \& {Bullock}, J.~S. 2006, ApJ,
  submitted, astro-ph/0612228

\bibitem[{{Seigar} {et~al.}(2006){Seigar}, {Graham}, \&
  {Jerjen}}]{seigar06}
{Seigar}, M.~S., {Graham}, A.~W., \& {Jerjen}, H. 2006, MNRAS,
  submitted, astro-ph/0612229

\bibitem[{{Siegel} {et~al.}(2002){Siegel}, {Majewski}, {Reid}, \&
  {Thompson}}]{Siegel_etal02}
{Siegel}, M.~H., {Majewski}, S.~R., {Reid}, I.~N., \& {Thompson}, I.~B. 2002,
  ApJ, 578, 151

\bibitem[{{Somerville} \& {Kolatt}(1999)}]{Somerville99}
{Somerville}, R.~S. \& {Kolatt}, T.~S. 1999, MNRAS, 305, 1

\bibitem[{{Somerville} \& {Primack}(1999)}]{sp:99}
{Somerville}, R.~S. \& {Primack}, J.~R. 1999, MNRAS, 310, 1087

\bibitem[{{Sommer-Larsen}(2006)}]{Sommer-Larsen06}
{Sommer-Larsen}, J. 2006, MNRAS, 369, 958

\bibitem[{{Stewart} {et~al.}(2008)}]{Stewart_etal08}
{Stewart}, K., {Bullock}, J.~S., {Wechsler}, R.~H.,{Maller}, A.~H., \& {Zentner}, A.~R. 2008, ApJ, accepted, astro-ph/0711.5027

\bibitem[{{Tinker} {et~al.}(2005){Tinker}, {Weinberg}, {Zheng}, \&
  {Zehavi}}]{tinker:05}
{Tinker}, J.~L., {Weinberg}, D.~H., {Zheng}, Z., \& {Zehavi}, I. 2005, ApJ,
  631, 41

\bibitem[{{Tissera} {et~al.}(2005){Tissera}, {De Rossi}, \&
  {Scannapieco}}]{Tissera_etal05}
{Tissera}, P.~B., {De Rossi}, M.~E., \& {Scannapieco}, C. 2005, MNRAS, 364,
  L38

\bibitem[{{Tremonti} {et~al.}(2004){Tremonti}, {Heckman}, {Kauffmann},
  {Brinchmann}, {Charlot}, {White}, {Seibert}, {Peng}, {Schlegel}, {Uomoto},
  {Fukugita}, \& {Brinkmann}}]{Tremonti_etal04}
{Tremonti}, C.~A., {Heckman}, T.~M., {Kauffmann}, G., {Brinchmann}, J.,
  {Charlot}, S., {White}, S.~D.~M., {Seibert}, M., {Peng}, E.~W., {Schlegel},
  D.~J., {Uomoto}, A., {Fukugita}, M., \& {Brinkmann}, J. 2004, ApJ, 613, 898

\bibitem[{{van den Bosch} {et~al.}(2007){van den Bosch}, {Yang}, {Mo},
  {Weinmann}, {Macci{\`o}}, {More}, {Cacciato}, {Skibba}, \& {Kang}}]{vdb07}
{van den Bosch}, F.~C., {Yang}, X., {Mo}, H.~J., {Weinmann}, S.~M.,
  {Macci{\`o}}, A.~V., {More}, S., {Cacciato}, M., {Skibba}, R., \& {Kang}, X.
  2007, MNRAS, 376, 841

\bibitem[{{Venn} {et~al.}(2004){Venn}, {Irwin}, {Shetrone}, {Tout}, {Hill}, \&
  {Tolstoy}}]{venn_etal04}
{Venn}, K.~A., {Irwin}, M., {Shetrone}, M.~D., {Tout}, C.~A., {Hill}, V., \&
  {Tolstoy}, E. 2004, AJ, 128, 1177

\bibitem[{{White} \& {Rees}(1978)}]{White78}
{White}, S.~D.~M. \& {Rees}, M.~J. 1978, MNRAS, 183, 341

\bibitem[{{Williams} {et~al.}(2007){Williams}, {Ciardullo}, {Durrell},
  {Vinciguerra}, {Feldmeier}, {Jacoby}, {Sigurdsson}, {von Hippel}, {Ferguson},
  {Tanvir}, {Arnaboldi}, {Gerhard}, {Aguerri}, \& {Freeman}}]{Williams_etal07}
{Williams}, B.~F., {Ciardullo}, R., {Durrell}, P.~R., {Vinciguerra}, M.,
  {Feldmeier}, J.~J., {Jacoby}, G.~H., {Sigurdsson}, S., {von Hippel}, T.,
  {Ferguson}, H.~C., {Tanvir}, N.~R., {Arnaboldi}, M., {Gerhard}, O.,
  {Aguerri}, J.~A.~L., \& {Freeman}, K. 2007, ApJ, 656, 756

\bibitem[{{Willman} {et~al.}(2004){Willman}, {Governato}, {Wadsley}, \&
  {Quinn}}]{Willman_etal04}
{Willman}, B., {Governato}, F., {Wadsley}, J., \& {Quinn}, T. 2004, MNRAS,
  355, 159

\bibitem[{{Yang} {et~al.}(2003){Yang}, {Mo}, \& {van den Bosch}}]{yang_etal03}
{Yang}, X., {Mo}, H.~J., \& {van den Bosch}, F.~C. 2003, MNRAS, 339, 1057

\bibitem[{{Yanny} {et~al.}(2000){Yanny}, {Newberg}, {Kent},
  {Laurent-Muehleisen}, {Pier}, {Richards}, {Stoughton}, {Anderson}, {Annis},
  {Brinkmann}, {Chen}, {Csabai}, {Doi}, {Fukugita}, {Hennessy}, {Ivezi{\'c}},
  {Knapp}, {Lupton}, {Munn}, {Nash}, {Rockosi}, {Schneider}, {Smith}, \&
  {York}}]{Yanny_etal00}
{Yanny}, B., {Newberg}, H.~J., {Kent}, S., {Laurent-Muehleisen}, S.~A., {Pier},
  J.~R., {Richards}, G.~T., {Stoughton}, C., {Anderson}, Jr., J.~E., {Annis},
  J., {Brinkmann}, J., {Chen}, B., {Csabai}, I., {Doi}, M., {Fukugita}, M.,
  {Hennessy}, G.~S., {Ivezi{\'c}}, {\v Z}., {Knapp}, G.~R., {Lupton}, R.,
  {Munn}, J.~A., {Nash}, T., {Rockosi}, C.~M., {Schneider}, D.~P., {Smith},
  J.~A., \& {York}, D.~G. 2000, ApJ, 540, 825

\bibitem[{{Zaritsky} {et~al.}(1994){Zaritsky}, {Kennicutt}, \&
  {Huchra}}]{Zaritsky_etal94}
{Zaritsky}, D., {Kennicutt}, Jr., R.~C., \& {Huchra}, J.~P. 1994, ApJ, 420, 87

\bibitem[{{Zentner}(2007)}]{zentner06}
{Zentner}, A.~R. 2006, Int. J. Mod. Phys. D, 16, 763, [arXiv:astro-ph/0611454]

\bibitem[{{Zentner} {et~al.}(2005){Zentner}, {Berlind}, {Bullock}, {Kravtsov},
  \& {Wechsler}}]{Zentner05}
{Zentner}, A.~R., {Berlind}, A.~A., {Bullock}, J.~S., {Kravtsov}, A.~V., \&
  {Wechsler}, R.~H. 2005, ApJ, 624, 505

\bibitem[{{Zentner} \& {Bullock}(2003)}]{zb:03}
{Zentner}, A.~R. \& {Bullock}, J.~S. 2003, ApJ, 598, 49

\bibitem[{{Zibetti} {et~al.}(2005){Zibetti}, {White}, {Schneider}, \&
  {Brinkmann}}]{Zibetti05}
{Zibetti}, S., {White}, S.~D.~M., {Schneider}, D.~P., \& {Brinkmann}, J. 2005,
  MNRAS, 358, 949

\end{thebibliography}
\end{document}